\documentclass[reprint,superscriptaddress,nofootinbib,amsmath,amssymb,aps,floatfix]{revtex4-1}
\pdfoutput=1
\usepackage{graphicx}
\usepackage{dcolumn}
\usepackage{bm}
\usepackage[utf8]{inputenc}
\usepackage{graphics,graphicx}
\usepackage{parskip}
\usepackage{amsmath,amssymb}
\usepackage[normalem]{ulem}
\usepackage{amsfonts}

\usepackage{multirow}
\usepackage{fancyhdr}
\usepackage{xcolor}
\usepackage{placeins}
\usepackage[title]{appendix}
\usepackage{wasysym}
\usepackage{url}
\usepackage{braket}
\usepackage{siunitx}
\usepackage{soulpos}
\usepackage{float}
\usepackage[caption=false]{subfig}

\usepackage{lipsum}
\usepackage{booktabs}
\usepackage{comment}
\usepackage{braket}
\usepackage[colorlinks=true,filecolor=.]{hyperref}
\usepackage{cleveref}
\usepackage{natbib}
\usepackage{setspace}
\usepackage{subfiles} 

\begin{document}

\title{Excited state preparation of trapped ultracold atoms via swept potentials}

\author{Daniel J. Bosworth} 
 \email{dboswort@physnet.uni-hamburg.de}
\affiliation{%
 Zentrum f\"ur Optische Quantentechnologien, Universit\"at Hamburg,\\ Luruper Chaussee 149, 22761 Hamburg, Germany\\
}%
\affiliation{%
 The Hamburg Centre for Ultrafast Imaging, Universit\"at Hamburg,\\ Luruper Chaussee 149, 22761 Hamburg, Germany\\
}%

\author{Maxim Pyzh}%
\email{mpyzh@physnet.uni-hamburg.de}
\affiliation{%
 Zentrum f\"ur Optische Quantentechnologien, Universit\"at Hamburg,\\ Luruper Chaussee 149, 22761 Hamburg, Germany\\
}%

\author{Peter Schmelcher}
\email{pschmelc@physnet.uni-hamburg.de}
\affiliation{%
 Zentrum f\"ur Optische Quantentechnologien, Universit\"at Hamburg,\\ Luruper Chaussee 149, 22761 Hamburg, Germany\\
}%
\affiliation{%
 The Hamburg Centre for Ultrafast Imaging, Universit\"at Hamburg,\\ Luruper Chaussee 149, 22761 Hamburg, Germany\\
}%

\date{\today}

\begin{abstract}
We study the out-of-equilibrium dynamics of non-interacting atoms confined within a one-dimensional harmonic trap triggered by dragging an external long-range potential through the system. The symmetry-breaking nature of this moving potential couples adjacent eigenstates in the atoms' effective potential, leading to an energy landscape reminscent of systems exhibiting trap-induced shape resonances. These couplings may be exploited to selectively excite the atoms into higher vibrational states of the harmonic trap by controlling the motion of the dragged potential. To this end, we consider two protocols designs: the first protocol strives to maintain adiabaticity at critical points during the atoms' dynamics, whilst the second protocol utilises the fast tunnelling of the atoms within their effective double-well potential. These protocols take place in the few to many millisecond regime and achieve high-fidelity excitation of the atoms into pure vibrational states and superpositions thereof. Overall, our study highlights the significance of dragged potentials for controlling and manipulating atom dynamics and offers intuitive protocols for achieving desired excitations.
\end{abstract}

\maketitle

\section{Introduction}
Reliable and efficient quantum state engineering techniques are indispensable for emerging quantum technologies from information processing to interferometry and communications. Ultracold quantum gases are particularly suited to the manipulation of quantum states and dynamics due to the exceptional control over inter-particle interactions via tunable scattering resonances~\cite{Chin2010Feshbach}, the ability to prepare ensembles with a well-defined number of particles~\cite{Kaufman2014Twoparticle,Serwane2011Deterministic} and the flexibility of trapping geometry in terms of shape~\cite{Henderson2009Experimental,Morizot2006Ring}, periodicity~\cite{Bloch2005Ultracold} and dimensionality~\cite{Petrov2004Lowdimensional}.\\
For trapped ultracold species, transitions between different vibrational states can be carried out by employing external drives, such as deforming~\cite{Martinez-Garaot2013Vibrational} or shaking the trapping potential~\cite{Soerensen2018Quantum}. The latter approach was implemented in~\cite{Buecker2011Twin} to transfer a BEC to the collective first excited trap state, which serves as a twin-beam matter wave source upon collisional de-excitation of the atoms. Similar protocols for population-inversion have been proposed which rely on adiabatic cycles controlled by the interaction of the BEC with a $\delta$-like impurity~\cite{Tanaka2016Complete,Tanaka2020Generating}. Additionally, collective excitations such as solitons and vortices may be generated through appropriately steering a focused laser beam through a condensate~\cite{Madison2000Vortex,Karkuszewski2001Method,Damski2001Simple,Damski2002Stirring,Hans2015Generating}. State transfer protocols such as those described above have been further combined with sophisticated quantum optimal control (QOC) techniques~\cite{Koch2022Quantum} and shortcuts to adiabaticity (STA)~\cite{Deffner2017Quantum,Guery-Odelin2019Shortcuts} in order to manipulate quantum systems with high-fidelity on timescales shorter than decoherence times (see e.g.~\cite{Bucker2013Vibrational}). QOC and STA can be used for a wide range of applications, including for example the transport of trapped ions~\cite{Bowler2012Coherent,Walther2012Controlling}.\\
Controlled collisions between species in separate traps provide a further avenue for quantum state engineering. In their theoretical work~\cite{Stock2003Quantum}, Stock \textit{et al.} found that the quantised relative motion of a colliding atom pair leads to resonances between trap eigenstates and molecular bound states which would not be present in free space. They termed these `trap-induced shape resonances' (TISR). Later theoretical works uncovered TISR for colliding atom-ion pairs~\cite{Idziaszek2007Controlled} and proposed using TISR to realise two-qubit quantum gates~\cite{Doerk2010Atomion} and excite atoms into higher Bloch bands of an optical lattice~\cite{Krych2009Controlled}. TISR have also been considered in the context of single atoms interacting with multiple impurities~\cite{Sroczynska2018Trapinduced}. Recently, a landmark experiment by Ruttley \textit{et al.}~\cite{Ruttley2023Formation} demonstrated the ‘mergoassociation’ of single cold RbCs molecules using TISR between the constituent atoms confined in separate optical tweezers.\\
In this work, we consider a particular case of an external drive which enables fine control over the vibrational state occupation of trapped atoms. Similar to~\cite{Karkuszewski2001Method,Damski2001Simple,Damski2002Stirring}, the external drive takes the form of dynamically-swept external potentials. Our potential however is repulsive at short range with a long-range attractive tail supporting bound states, which offers additional flexibility in terms of protocol design and a more diverse dynamical response of the system.~We explore how tuning the shape and drag speed of the external drive can be exploited to excite ground-state atoms into excited trap states or superpositions thereof. We propose two different types of protocols for achieving state transfer which rely on avoided crossings arising in the atoms' discrete energy spectrum due to the swept potential, in a manner analogous to the emergence of TISR. The first protocol, slow yet robust, relies on adiabatic sweeping of the potential around certain critical avoided crossings in the energy spectrum. The second protocol, significantly faster yet requiring precise control over the external potential's position, exploits the ability of the atom to undergo relatively fast tunnelling at the avoided crossings.\\
Our work is laid out as follows. In Section~\ref{sec:results-1}, we introduce the setup and discuss the landscape of avoided crossings arising in our system and how these can be used to shuttle the atoms to higher excited states. Section~\ref{sec:results-2} and section~\ref{sec:results-3} focus on the two different state preparation protocols and include proof-of-principle demonstrations for both as well as a discussion of their limitations. Section~\ref{sec:summary} summarises the present study and discusses directions for future work.\\
\section{Swept potential model}~\label{sec:results-1}
\begin{figure}[t]
    \centering
    \includegraphics[width = 0.5\textwidth]{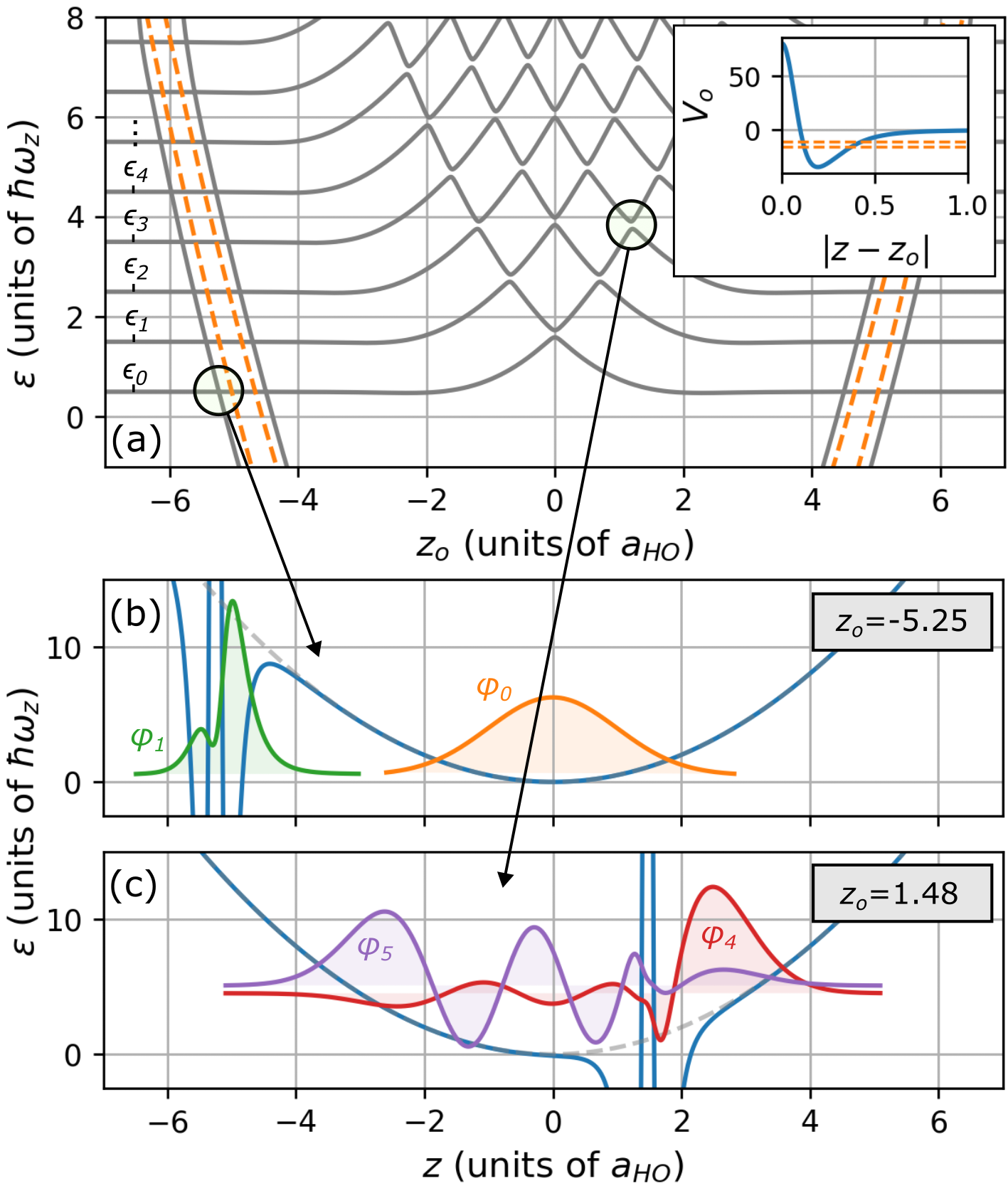}
    \caption{\textbf{Instantaneous single-particle energy spectrum.} (a) Discrete atomic energy spectrum as a function of the position $z_o$ of the external potential (see inset) relative to the trap centre. The energies of the lowest few harmonic trap eigenstates are labelled $\epsilon_i$. The dashed lines show the approximate energy shift of the external potential's bound states $\bar{\epsilon}_i + z_o^2/2$~\cite{Idziaszek2007Controlled} as a function of $z_o$, where $\bar{\epsilon}_i$ are the energies of the bound states without the harmonic trap. The solid circles highlight examples of narrowly-avoided crossings. (b) and (c) show plots of the instantaneous eigenstates $\{\varphi_i(z;z_o)\}$ near an avoided crossing between (b) a bound state of the dragged potential and a trap state and (c) two trap states. The solid blue line shows the effective potential experienced by the atoms and the dashed gray line is the harmonic trap potential. Eigenstates are vertically offset by their energy. Here, we have used the parameters $a = 120$, $b = 4\sqrt{10~c}$ and $c = 40$ for the external potential~\eqref{eq:potential}.}
    \label{fig:fig1}
\end{figure}
\begin{figure}[t]
    \centering
    \includegraphics[width = 0.5\textwidth]{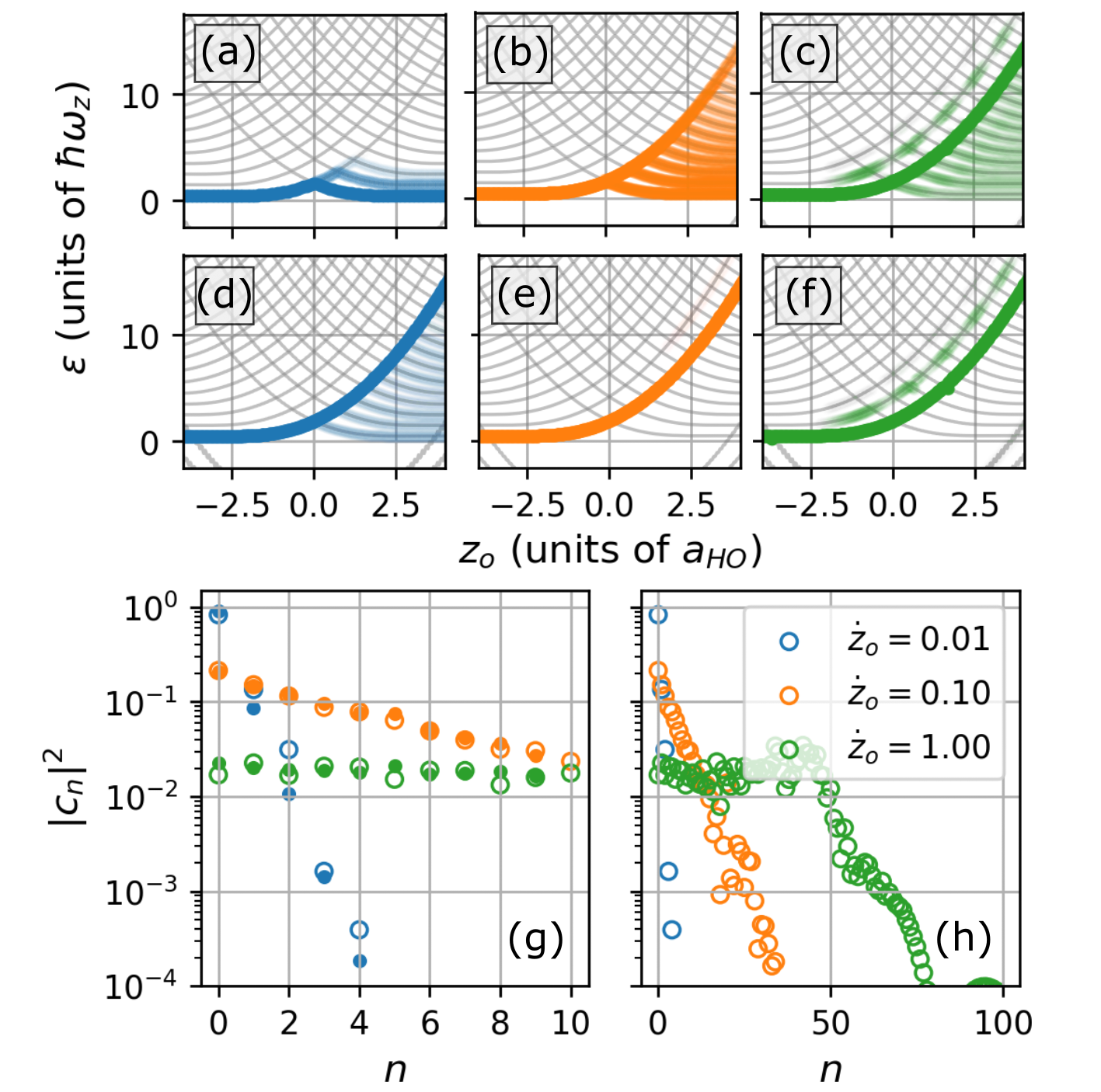}
    \caption{\textbf{Path of atomic state along the energy curves.} (a)-(f) The atomic energy spectrum (grey) weighted by the overlap of the atomic state with the instantaneous eigenstates $|\braket{\psi(z;z_o|\varphi_{n}(z;z_0)}|^2$ of the Hamiltonian~\eqref{eq:hamiltonian} as a function of $z_o(t)$ for constant drag speeds (a) $\dot{z}_o = 0.01$, (b) $\dot{z}_o = 0.10$ and (c) $\dot{z}_o = 1.00$. The model parameters are the same as those given in Fig.~\ref{fig:fig1}. (d)-(f) Shows the same as (a)-(c) for a barrier height $a=320$. (g),(h) Open circles show the overlap of the final state $|\braket{\psi_f(z)|\phi_{n}(z)}|^2$ with the first (g) 11 (h) 101 harmonic trap eigenstates $\{\phi_{n}\}$ ($n = 0,1,2,\ldots$) for various drag speeds $\dot{z}_o$. Filled circles in (g) are values obtained using the Landau-Zener formula~\eqref{eq:landau-zener}.}
    \label{fig:fig2}
\end{figure}
We begin in section~\ref{ssec:background} by introducing the time-dependent Hamiltonian which models the collision between a dragged external potential and trapped atoms in one spatial dimension. Section~\ref{ssec:constant-speed} considers the scenario in which the external potential is swept through the trap at a constant velocity, highlighting the emergence of avoided crossings between eigenstates during the collision and the role played by the potential's profile and speed. This motivates the discussion of the state preparation protocols which are the focus of Sections~\ref{sec:results-2} and~\ref{sec:results-3}.
\subsection{Model: collision of trapped atoms with a swept potential}~\label{ssec:background}
Our system is comprised of atoms of mass $m$ confined within a quasi-1D harmonic trap centred at the origin. The quasi-1D confinement requires $\omega_{\parallel}\ll\omega_{\perp}$, where $\omega_{\parallel}$ and $\omega_{\perp}$ are the longitudinal and transverse trapping frequencies, respectively. The longitudinal axis is chosen to be parallel with the $z$-axis and the corresponding longitudinal eigenstates and associated eigenenergies are denoted by $\{\phi_n(z)\}$ and $\{\epsilon_n = n+1/2\},\,\,n \in \mathbb{N}$. Transverse excitations are neglected throughout this paper, such that we restrict ourselves to a one-dimensional problem. Finally, unless stated otherwise all quantities are given in units defined by the oscillator length $a_{\text{HO}} = \sqrt{\hbar/m\omega_z}$ and the energy spacing $\varepsilon_{\text{HO}} = \hbar \omega_z$ of the longitudinal eigenstates.\\
At $t=0$, the atom occupies the trap's vibrational ground state $\phi_0(z)$. For $t > 0$, it experiences an additional time-dependent potential $V_o(z,t)$ which is swept from one side of the system to the other along the $z$-axis. The dragged potential's profile is comprised of a short-range repulsive barrier with an attractive long-range tail and takes the form
\begin{equation}
    \label{eq:potential}
    V_o(z,z_o(t)) = a e ^{-b (z-z_o(t))^2} - \frac{1}{2(z-z_o(t))^4 + 1/c},
\end{equation}
where $z_o(t)$ is the displacement of the repulsive barrier from the centre of the trap. The model parameters $a$, $b$ and $c$ set the height and width of the barrier as well as the depth of the wells formed by the attractive tail, respectively. A plot of the potential is provided in the inset of Fig.~\ref{fig:fig1} (a). This potential could be created in an experiment using, for example, a tightly-trapped ion~\cite{Schurer2014Groundstate,Schurer2015Capture,Tomza2019Cold} or a shaped optical potential~\cite{Henderson2009Experimental}.\\
Summarising the above considerations, we write the single atom Hamiltonian as
\begin{equation}
    \label{eq:hamiltonian}
        \hat{H}(z_o(t)) = -\frac{1}{2}\frac{d^2}{d z^2} + V_{\text{trap}}(z) + V_o(z,z_o(t)),
\end{equation}
where $V_{\text{trap}}(z) = z^2/2$ describes the time-independent harmonic trap. Eq.~\eqref{eq:hamiltonian} is parametrically-dependent on the position of the dragged potential $z_o$ and we denote its eigenstates and eigenvalues with $\{\varphi_{n}(z;z_o)\}$ and $\{\varepsilon_{n}(z_o)\}$, respectively, to contrast them with those of the pure harmonic trap ($\phi_n(z)$ and $\epsilon_n$). \\
Hamiltonians of the above form were employed in related works to model atom-ion interactions in the ultracold regime~\cite{Schurer2014Groundstate,Schurer2015Capture,Schurer2016Impact,Schurer2017Unraveling,Bosworth2021Spectral}. The atom-ion interaction has a species-dependent length scale $R^*$ and together with the harmonic oscillator length $a_{\text{HO}}$, these constitute the two characteristic length scales of the system. In this and prior works, we have been interested in the regime where these length scales are comparable. The form of our Hamiltonian in Eq.~\eqref{eq:hamiltonian} is valid for $R^* = a_{\text{HO}}$, which is valid for a variety of species in terms of the atom-ion interaction range and the achievable trapping frequencies. The analysis carried out in the remainder of this work holds for $R^* = a_{\text{HO}}$. However, we emphasise that the proposed protocols would also work for different values, so long as the length scales remain comparable. As we mentioned above, the model potential could be realised in one of two ways: (i) a trapped ion (ii) a shaped optical potential. Thus, we see that the realisation with a trapped ion is less flexible since the length scale $R^*$ is set by the choice of species for the atom-ion pair. In contrast, an optical potential would allow greater flexibility since the size of the potential may be tuned in addition to the trapping frequency. 
\subsection{Impact of swept potential on the atomic energy spectrum}~\label{ssec:constant-speed}
Let us first solve the time-independent problem to clarify the $z_o$-dependence of the atoms' discrete energy spectrum $\{\varepsilon_{n}(z_o)\}$. We choose the following model parameters for the external potential~\eqref{eq:potential}: $a = 120$, $b = 4\sqrt{10~c}$ and $c = 40$ (as used in ~\cite{Schurer2014Groundstate,Schurer2015Capture,Schurer2016Impact,Schurer2017Unraveling,Bosworth2021Spectral}). For this choice of parameters, the potential supports two bound states with energies $\bar{\epsilon}_0 = -12.2$ and $\bar{\epsilon}_1 = -10.4$, shown in the inset of Fig.~\ref{fig:fig1}~(a).\\
Fig.~\ref{fig:fig1}~(a) shows the evolution of the lowest nine eigenvalues with $z_o$, obtained using exact diagonalisation of the Hamiltonian~\eqref{eq:hamiltonian}. For $|z_o| > 6$, the lowest eigenstates have a regular energy-spacing $\hbar\omega_z$ and describe states of the unperturbed harmonic trap. Closer to the trap centre ($4 < |z_o(t)| < 6$), the energies of the external potential's bound states are reduced which leads to level-repulsions between the bound states and the trap eigenstates, generating two chains of avoided crossings. The avoided crossings seen here can be considered analogous to \textit{trap-induced shape resonances} (TISR), first predicted by Stock \textit{et al.} for colliding pairs of trapped atoms~\cite{Stock2003Quantum}. That these are indeed a form of shape resonance can be seen in Fig.~\ref{fig:fig1}~(b), which shows the trap's ground-state near its avoided crossing with the lower bound state of the external potential at $z_o = -5.25$. Here, these near-degenerate eigenstates are separated by a barrier that forms in the atom's effective potential created by the sum of $V_o(z,z_o)$ and $V_{\text{trap}}(z)$. In addition, a second variety of TISR-analogues manifests in this system due this time to the short-range repulsive barrier component of Eq.~\eqref{eq:potential}. One such example is shown in Fig.~\ref{fig:fig1}~(c), where two (perturbed) trap states are separated on either side of the external potential's Gaussian barrier at $z_o = 1.48$. We see therefore that the repulsive and attractive components of~\eqref{eq:potential} each create their own class of avoided crossings. Crucially, both kinds of shape resonances present in Fig.~\ref{fig:fig1} would not appear in the absence of the trap's discrete energy spectrum.\\
Let us now turn to the time-dependent solution of the Hamiltonian~\eqref{eq:hamiltonian}. In the remainder of this section, we examine the simplest case of the external potential Eq.~\eqref{eq:potential} moving at a constant velocity $\dot{z_o}$ from one side of the system to the other. We are interested in the state of the atoms at long times, i.e. after the external potential has passed into and through the system and excited it on the other side, and which factors influence it.\\
At $t=0$, the atoms occupy the ground-state of the trap $\varphi(z,0) = \phi_0(z)$. We choose the same model parameters for the external potential as before. For numerical purposes, we set the external potential's position at $t=0$ to be $z_o(0) = -6$, which is sufficiently far-removed from the trap centre to prevent an immediate quench of the initial atomic state. We determine the atomic dynamics $\psi = \psi(t)$ by solving the time-depdendent Schr\"{o}dinger equation via wavepacket propagation using a dynamically-optimised truncated basis representation~\cite{Cao2017Unified}.\\ 
We first consider the way in which the dragged potential couples the initial atomic state with other eigenstates during the course of the dynamics. For this purpose, we determine the overlap of the atomic state with the instantaneous eigenstates of the Hamiltonian~\eqref{eq:hamiltonian} as a function of $z_o(t)$. Fig.~\ref{fig:fig2}~(a)-(f) show plots of the energy spectrum (cf. Fig.~\ref{fig:fig1}~(a)) in which the curves $\{\varepsilon_{n}(z_o)\}$ are weighted by the overlap integrals $|\braket{\psi(t)|\varphi_n(z_o)}|^2$ for different drag speeds $\dot{z_o}$ and heights $a$ of the repulsive barrier. These plots effectively describe how $\psi(t)$ evolves within the Hilbert space of the Hamiltonian~\eqref{eq:hamiltonian}. We see in Fig.~\ref{fig:fig2}~(a) that for a sufficiently slow drag speed and small barrier height, the state $\psi(t)$ initially evolves along a single energy curve, with only minor population of neighbouring curves occuring after the dragged potential passes through the trap centre. For faster drag speeds and a greater barrier height, the atomic state follows an increasingly diabatic path to higher energy curves. Fig.~\ref{fig:fig2} shows that for $\dot{z_o} = 0.01$ and $\dot{z_o} = 0.10$, diabatic transitions between energy curves take place exclusively at the avoided crossings, since there the coupling between energy curves is greatest and the energy gap smallest. However, this simple picture breaks down at sufficiently fast drag speeds, such as at $\dot{z_o} = 1.00$ which is shown in Fig.~\ref{fig:fig2} (c) and (f). In both of these cases, the coupling between curves becomes strong enough that additional transitions take place at positions $z_o$ away from the immediate vicinity of the avoided crossings, where the curves have relatively large energy separations. For our purposes, these additional transitions are undesirable since they constitute an additional form of `leakage' between energy curves which hinders the controlled preparation of a well-defined final atomic state.\\
A more quantitative understanding of the influence of the drag speed and barrier height on the path of the atomic state in Fig.~\ref{fig:fig2} is provided by the semi-classical Landau-Zener formula~\cite{Landau1932Zur,Zener1932NonadiabaticA}. This determines the probability $P_{ij}$ for a diabatic transition at an avoided crossing between the energy curves of the eigenstates $\varphi_i(z_o)$ and $\varphi_j(z_o)$: 
\begin{equation}
    \label{eq:landau-zener}
    P_{ij} = \exp \bigg(-2\pi\frac{\Delta^2_{ij}}{\dot{z}_o\,\alpha_{ij}}\bigg).
\end{equation}
Here, $\Delta_{ij} = \text{min}(|\varepsilon_i-\varepsilon_j|)/2$ is half the minimum energy gap at the avoided crossing and $\alpha_{ij} = |\frac{d}{d z_o}(\varepsilon_i - \varepsilon_j)|$. For $P_{ij}\rightarrow 0$, transitions between the states are suppressed, i.e. the dynamics is adiabatic. This holds for the condition $\Delta^2_{ij}\gg \dot{z}_o \alpha_{ij}$. Whereas for $\Delta^2_{ij}\ll \dot{z}_o \alpha_{ij}$, $P_{ij}\rightarrow 1$ and the dynamics is maximally diabatic.\\
The filled circles in Fig.~\ref{fig:fig2}~(g) are predictions for the composition of the atomic state at long times determined by applying Eq.~\eqref{eq:landau-zener} at each crossing encountered by the state. The predictions are in good agreement with the results obtained from the solution of the time-dependent Schr\"{o}dinger equation (open circles) over a wide range of drag speeds $\dot{z}_o$. Thus, we see that the Landau-Zener formula~\eqref{eq:landau-zener} is a reasonable model for describing the state's path and we may use it to guide our intuition. Fig.~\ref{fig:fig2}~(h) extends the numerical results from Fig.~\ref{fig:fig2}~(g) up to the $100^{\text{th}}$ excited trap state, highlighting that it is in principle possible to populate arbitrarily-highly excited states using the dragged potential. In an experimental setting however, the finite depth of the trapping potential imposes an upper energy limit and any atoms excited beyond this threshold would be lost from the system. This loss could be exploited to our advantage in the following way. We may design a state preparation protocol in which any atoms that do not reach the desired final state are lost from the system, thereby maximising the fidelity with the target state at the cost of particle number uncertainty. This could be used to circumvent the limitations of the adiabatic state preparation protocol which is the focus of Section~\ref{sec:results-2}.\\
From Eq.~\eqref{eq:landau-zener}, we see that we have three knobs at our disposal for controlling the atoms' path through the energy curves $\{\varepsilon_{n}(z_o)\}$: $\Delta_{ij}$, $\alpha_{ij}$ and $\dot{z}_o$. The gap size $\Delta_{ij}$ at each avoided crossing is determined by the size of the barrier at the shape resonance since taller, wider barriers lead to more narrowly-avoided crossings. Therefore, we can control $\Delta_{ij}$ by tuning the model parameters in Eq.~\eqref{eq:potential} as well as the longitudinal trapping frequency $\omega_z$. These will also influence $\alpha_{ij}$, however the quadratic dependence of $\Delta_{ij}$ in Eq.~\eqref{eq:landau-zener} makes it a more sensitive and thus attractive control parameter. The speed of the dragged potential is also an attractive control parameter since it is a free parameter.\\
In the following sections, we develop protocols which exploit these control parameters in order to realise deterministic state preparation, such that the dragged potential shuttles the atoms into an excited trap state $\phi_n,\,n>0$ or a well-defined superposition of $N$ trap states $\sum_{n=0}^{N}c_n\phi_n$. We denote the target state by $\psi_t$ and the goal of the following sections is to maximise the fidelity measure $\mathcal{F} = |\braket{\psi|\psi_t}|^2$. We choose the following fixed set of model parameters: $a=320,\,b = 4\sqrt{10~c}$ and $c = 40$. In particular, we choose $a=320$, since from Fig.~\ref{fig:fig2}~(e) we see that for this barrier height - in combination with a drag speed of $z_o=0.1$ - the state's path is predominantly diabatic and transitions between energy curves are to a large extent `clean', by which we mean that the transitions occur chiefly at the avoided crossings and not, as is the case in Fig.~\ref{fig:fig2}~(e) and (f), also in-between avoided crossings. Both of these features are crucial for realising efficient, high-fidelity state preparation protocols.\\
\section{Adiabatic protocol}~\label{sec:results-2}
\begin{figure}[t]
    \centering
    \includegraphics[width = 0.5\textwidth]{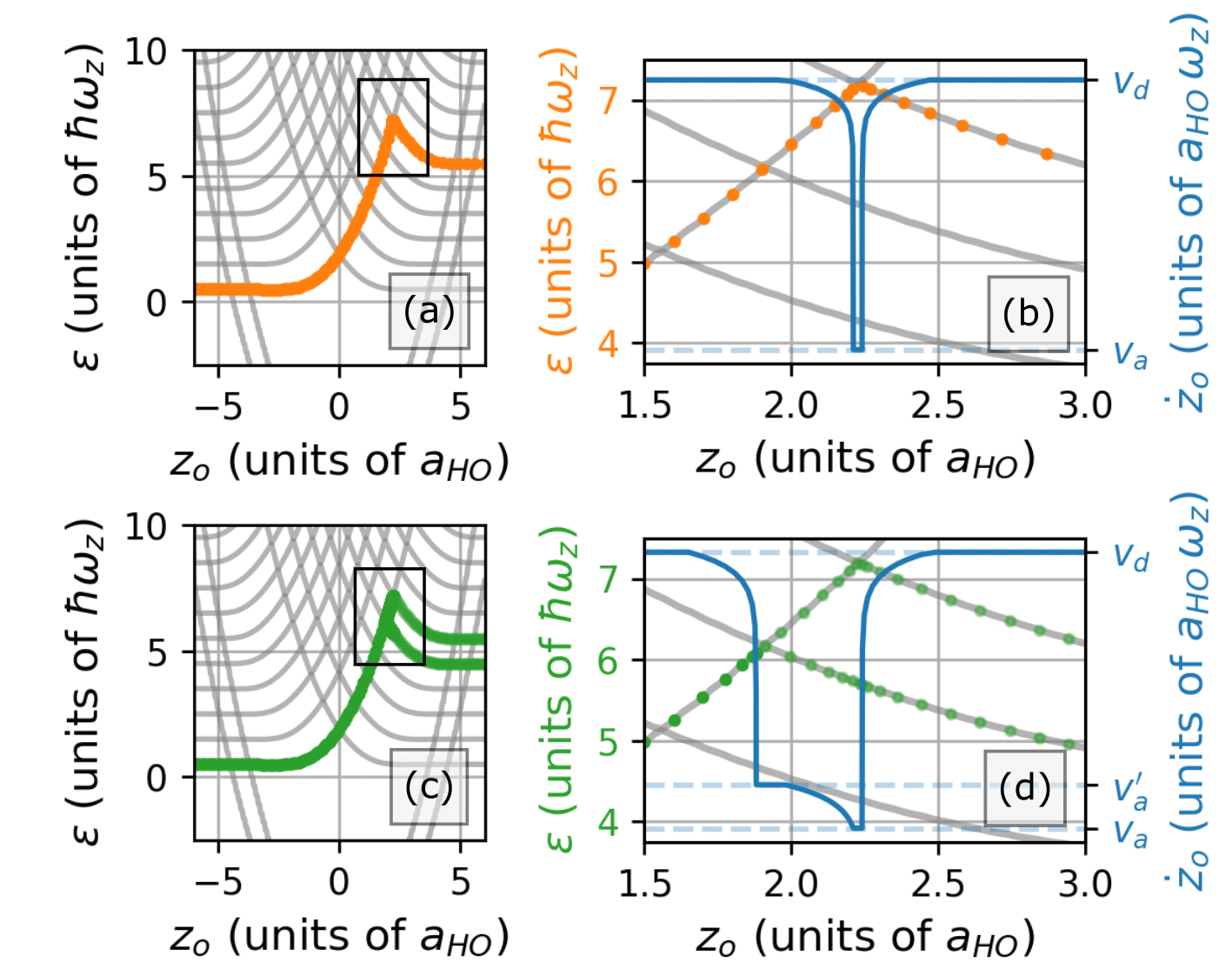}
    \caption[Adiabatic protocol schematic.]{\textbf{Schematic of the adiabatic protocol.} (a) The ideal state path (orange) through the atomic energy spectrum (grey) to excite the atom to the fifth excited trap state $\phi_5(z)$. (b) Close-up of the critical region highlighted by the box in (a). The impurity's drag speed is overlaid in blue, indicating the transition between the diabatic and adiabatic speeds ($v_d$ and $v_a$, respectively). (c) The ideal state path (green) through the atomic energy spectrum (grey) to excite the atom to the superposition state $(\phi_4+e^{i\Phi(t)}\phi_5)/\sqrt{2}$, where $\Phi(t) = -\omega_z t$. (d) Close-up of the critical region highlighted by the box in (c). Note that in both (b) and (d) $\dot{z}_o$ is plotted on a log-scale for the sake of visibility.}
    \label{fig:fig3}
\end{figure}
\begin{figure*}[t]
    \centering
    \includegraphics[width = \textwidth]{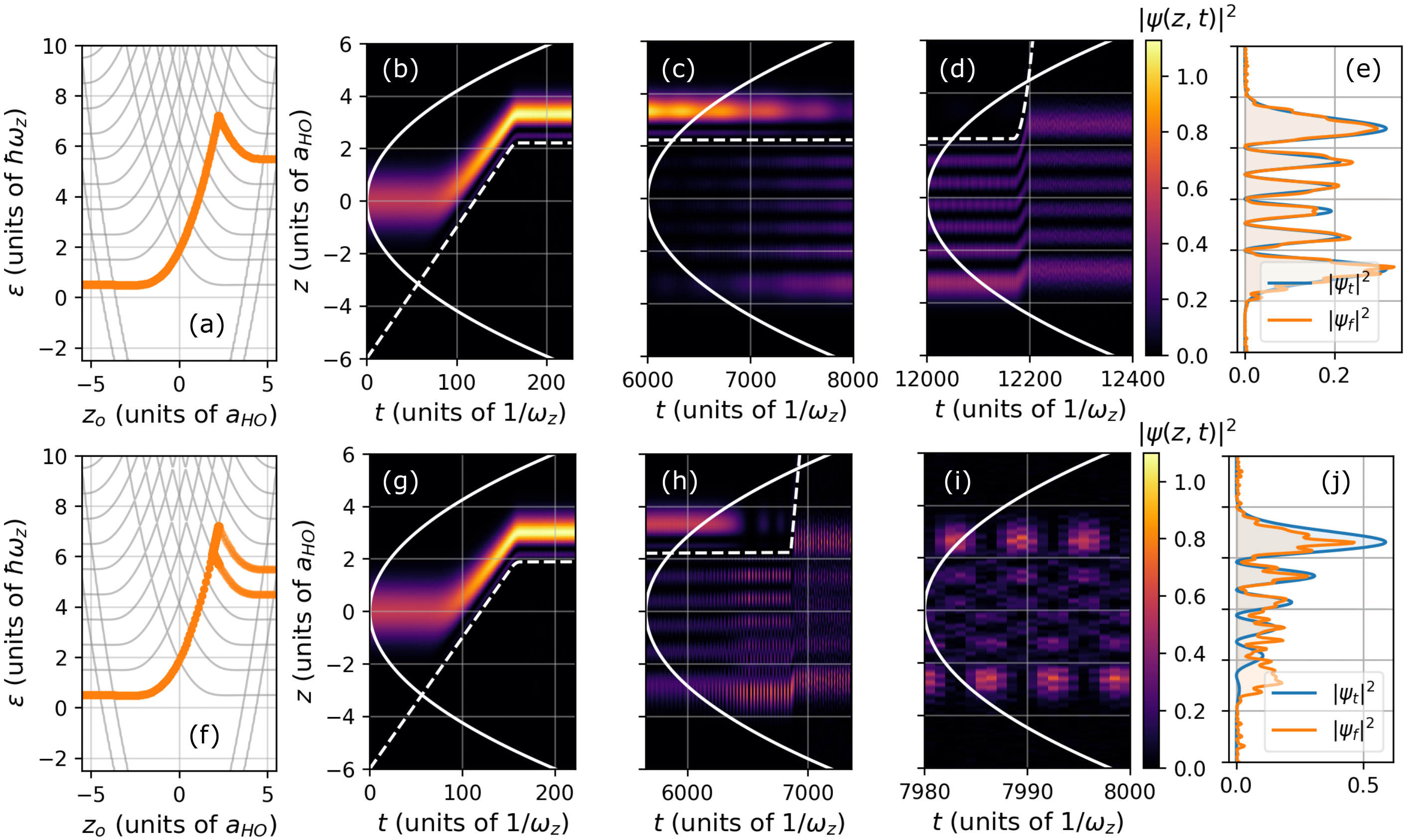}
    \caption[Adiabatic protocol results: proof-of-principle]{\textbf{Adiabatic protocol.} (a)-(e) Exciting atoms to the fifth excited trap state $\phi_5(z)$ using the adiabatic protocol. (a) The instantaneous energy spectrum (grey) with the coloured line representing the overlap of the atomic state with the instantaneous eigenstates $|\braket{\psi(z,z_o)|\varphi_{\nu}(z,z_o)}|^2$ of~\eqref{eq:hamiltonian} as a function of $z_o(t)$. (b)-(d) Atomic probability density $|\psi(z,t)|^2$ for different time intervals during the protocol. The solid lines represent the harmonic trap (scaled for visibility) and  dashed lines indicate $z_o(t)$. (e) Comparison of the density of the final atomic state $|\psi_f|^2$ to the target state $|\psi_t = \phi_5|^2$. Here, we achieve a fidelity $|\braket{\psi_f|\psi_{t}}|^2$ of 97.4\%. (f)-(g) Show the same as the top row for the target state $(\phi_4+e^{i\Phi}\phi_5)/\sqrt{2}$, where $\Phi(t) = -\omega_z t$. Here, we achieve a fidelity $|\braket{\psi_f|\psi_{t}}|^2$ of 92.6\%. For both protocols, $v_{d} = 0.1$. For the top row, $v_{a} = v_d/20,000$. For the bottom row, $v_{a} = v_d/600$ and $v_{a}^{\prime} = v_d/2000$ ($v_d$, $v_a$ and $v_{a}^{\prime}$ are defined in Fig.~\ref{fig:fig3}). Wavefunctions are normalised such that $\int dz |\psi(z)|^2 = 1$.}
    \label{fig:fig4}
\end{figure*}
\begin{table*}[t]
  \centering
  \begin{tabular}{|c|c|c|}
  \hline
     $v_a/v_d$ & $|\braket{\psi_f|\phi_5}|^2$ & $t_{\text{tot}}$ \\ \hline
    $5.0\times 10^{-3}$ & 25.8 & $3.20\times10^2$ \\ 
    $5.0\times 10^{-4}$ & 89.1 & $1.40\times10^3$  \\ 
    $5.0\times 10^{-5}$ & 97.4 & $1.22\times10^4$   \\ \bottomrule
    \hline
  \end{tabular}
  \caption[Results from adiabatic protocol for the fifth excited state.]{\textbf{Dependence of final fidelity on the adiabatic speed}. (I) Ratio of adiabatic $v_a$ and diabatic $v_d$ speeds, with $v_d = 0.1$ in all cases. (II) Fidelity with the target state (III) Protocol duration.}
  \label{table:adi-prot-results}
\end{table*}
This section introduces the first state preparation protocol, an adiabatic protocol, which seeks to control the path of the atomic state through the energy curves $\{\varepsilon_{n}(z_o)\}$ using only the intuition provided by the Landau-Zener model~\eqref{eq:landau-zener} discussed in Section~\ref{sec:results-1}. Specifically, we use the drag speed $\dot{z}_o$ of the external potential to control whether the state traverses a given crossing adiabatically or diabatically in order to force it to follow a pre-determined path through the energy spectrum. In particular, we demonstrate preparation of the target states $\psi_t^{(1)} = \phi_5$ and $\psi_t^{(2)}(t) = (\phi_4 + e^{i\Phi(t)}\phi_5)/\sqrt{2}$, where we include the phase factor $\Phi(t) = -\omega_z t$ to indicate that the latter target state is not a pure eigenstate of the harmonic trap and hence undergoes periodic dynamics. The demonstration of preparing a mixed state is used to highlight the versatility of the protocol. In principle however, it would be possible to employ similar protocols in order to engineer localised wavepackets in anharmonic trap potentials for probing quantum collapse and revival behaviour~\cite{Robinett2004Quantum}.\\
The adiabatic protocol is outlined in Fig.~\ref{fig:fig3}. In particular, Fig.~\ref{fig:fig3}~(a) illustrates the ideal path through the energy spectrum from the ground state to the fifth excited state of the harmonic trap $\phi_5$. Ten narrowly-avoided crossings lie along this particular path. Starting from $t=0$, the state should evolve diabatically at speed $v_{\text{d}}$ until just before it reaches the 8$^{\text{th}}$ avoided crossing (indicated by the box in Fig.~\ref{fig:fig3} (a)), whereupon the dragged potential is decelerated linearly to the speed $v_{\text{a}}$, which should be sufficiently slow to fulfill the adiabatic condition $\Delta^2_{ij}\gg \dot{z}_o \alpha_{ij}$ (see Fig.~\ref{fig:fig3}~(b)). If no deceleration occurs, the state will continue to populate higher trap eigenstates, similar to the path seen in Fig.~\ref{fig:fig2}~(e). After passing this critical 8$^{\text{th}}$ avoided crossing, the potential is accelerated once again to $v_{\text{d}}$ and the state continues diabatically through the last two avoided crossings, finally reaching the target state $\psi_t^{(1)} = \phi_5$.\\
Equally, the target state $\psi_t^{(2)}(t) = (\phi_4 + e^{i\Phi(t)}\phi_5)/\sqrt{2}$ may be achieved through a slight modification to the protocol for $\psi_t^{(1)}$. In particular, an additional deceleration step is required such that the state splits equally along the two energy curves at the 7$^{\text{th}}$ avoided crossing, as depicted in Fig.~\ref{fig:fig3}~(c). The speed protocol is shown in Fig.~\ref{fig:fig3}~(d). The potential is first decelerated from $v_d$ to $v_a^{\prime}$, whose value is chosen such that an equal mixing between states at the 7$^{\text{th}}$ crossing is achieved and can be estimated using Eq.~\eqref{eq:landau-zener}.\\
The results of the simulations for $\psi_t^{(1)}$ are summarised in the top row of Fig.~\ref{fig:fig4}. Fig.~\ref{fig:fig4}~(a) shows the actual path followed by the atomic state in each simulation, which agree as expected with the ideal path given in Fig.~\ref{fig:fig3}~(a). The evolution of the atomic probability density $\rho(z,t) = \psi^*(z,t)\psi(z,t)$ is provided in Fig.~\ref{fig:fig4}~(b)-(d) and the external potential's trajectory $z_o(t)$ is indicated by the dashed line. As the potential enters the trap (Fig.~\ref{fig:fig4}~(b)), the atomic density is swept in the direction of motion of the potential and the dynamics of the state is diabatic. After the external potential is decelerated, the density begins to tunnel to the opposite side of the potential's barrier Fig.~\ref{fig:fig4}~(c). As the potential leaves the trap (Fig.~\ref{fig:fig4}~(d)), the atomic density re-centres on $z=0$ and its profile matches approximately that of the fifth excited trap state (see comparison in Fig.~\ref{fig:fig4}~(e)). For this particular simulation, we obtain an overlap of 97.4\% with the target state in a time of $\sim1.22\times 10^4$.\\
Similar results for the $\psi_t^{(2)}$ protocol are depicted in the bottom row of Fig.~\ref{fig:fig4}. Here, we obtain an overlap of 92.6\% with the target state. The fidelity is smaller than that obtained for $\psi_t^{(1)}$ in part due to the larger value of $v_a$ used in this example (see Fig.~\ref{fig:fig4} caption). Consequently, the duration of this protocol is shorter at $\sim6.90\times 10^3$. The final atomic state exhibits regular density oscillations (Fig.~\ref{fig:fig4} (i)) with a period matching the time scale set by the energy separation of the neighbouring trap states, namely $2\pi/\omega_z$.\\
Adiabatic protocols are slow by nature. For a longitudinal trapping frequency of $\omega_z = 2\pi\times300~$Hz, the examples shown in Fig.~\ref{fig:fig4} would have a duration on the order of seconds. A tighter trapping potential would reduce this of course, since the time unit $\tau$ is given by $\tau = 1/\omega_z$ in our unit system. In addition, using larger values of $v_a$ would further reduce the protocol duration, but would come at the cost of the  fidelity (see Table~\ref{table:adi-prot-results}). Additional improvements could be made by minimising the distance over which the potential moves adiabatically via standard optimisation techniques.\\
The final fidelity achieved is strongly influenced by the value of $v_a$. Nonetheless, there are additional sources of fidelity loss, accounting overall for $\sim1\%$ of the total probability. Firstly, the state's evolution whilst the potential is dragged at $v_d$ is not perfectly diabatic, which leads to minor losses at each crossing. Diabatic transitions between energy curves away from the avoided crossings are a further source of loss, as we saw for fast drag speeds in Fig.~\ref{fig:fig2}~(c) and~(f). No doubt a protocol could be devised to fine-tune the drag speed around particular regions where these transitions become significant. This would, however, make the overall protocol more complex for rather marginal improvements to the fidelity.\\
\section{Tunnelling protocol}~\label{sec:results-3}
\begin{figure}[t]
    \centering
    \includegraphics[width=0.5\textwidth]{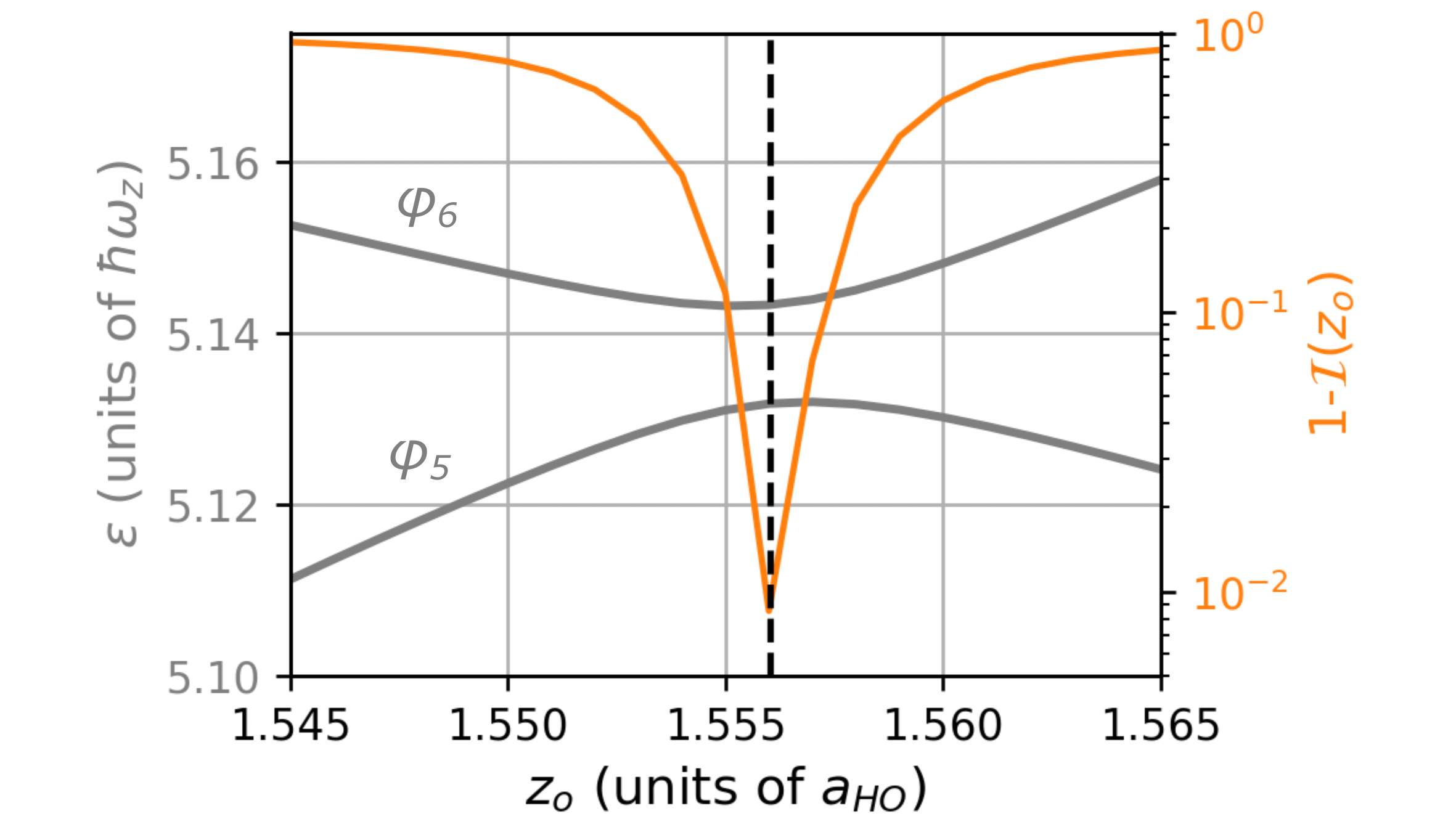}
    \caption[\textbf{Perfect tunnelling condition.}]{\textbf{Perfect tunnelling condition.} Energy curves $\varepsilon(z_o)$ for the eigenstates $\varphi_5(z;z_o)$ and $\varphi_6(z;z_o)$ in the vicinity of their avoided crossing (left axis). Also shown is $1-\mathcal{I}(z_o)$, where $\mathcal{I}(z_o)$ is the overlap integral $\mathcal{I}(z_o) = \int dz |\varphi_A(z,;z_o)||\varphi_B(z;z_o)|$ appearing in Eq.~\eqref{eq:integrals} (right axis).}
    \label{fig:tunnel-condition}
\end{figure}
\begin{figure*}[t]
    \centering
    \includegraphics[width=\textwidth]{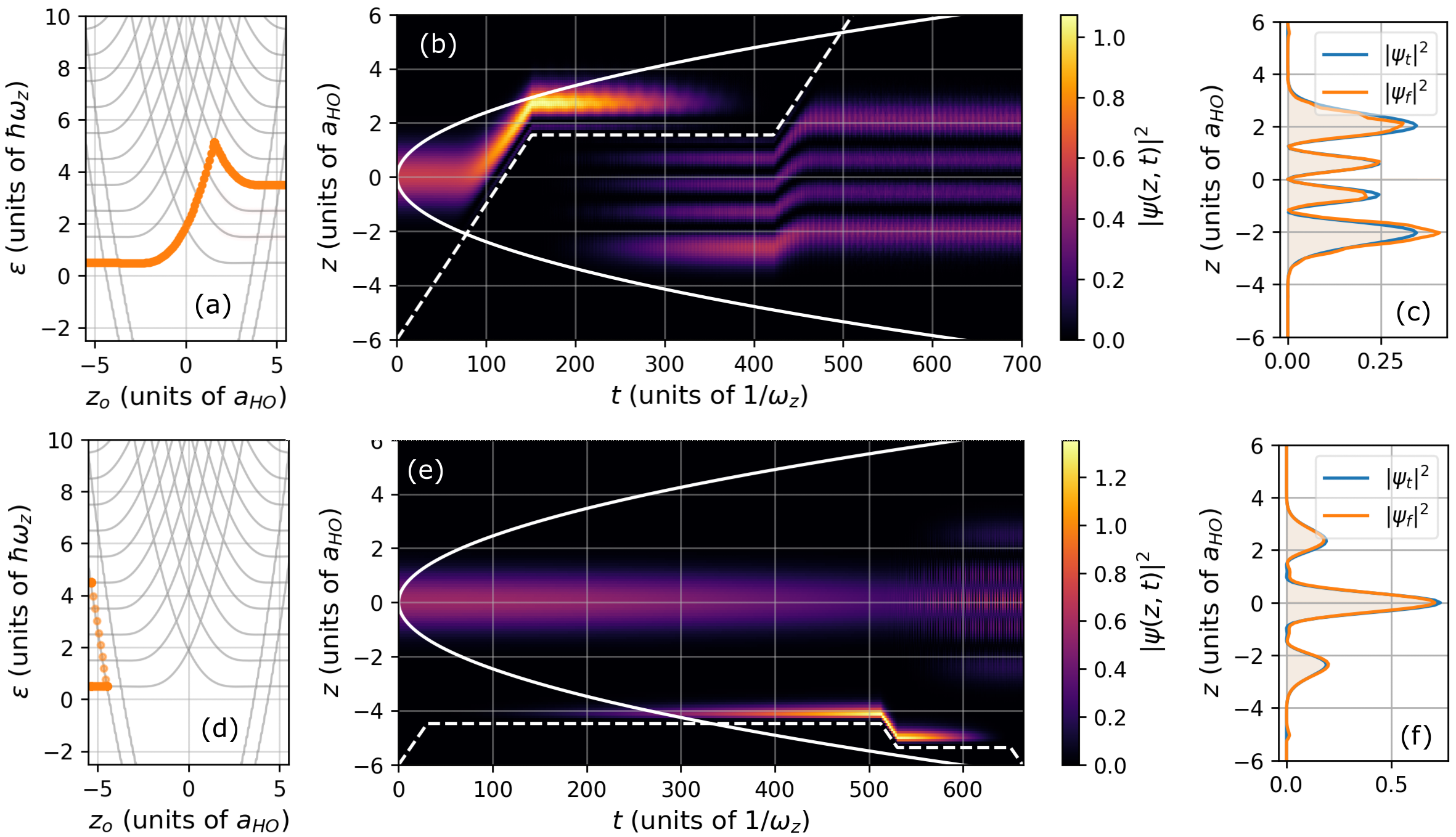}
    \caption[Tunnelling protocol results: proof-of-principle]{\textbf{Tunnelling protocol.} (a)-(c) Exciting atoms to the third excited trap state $\phi_3$ using the tunnelling protocol. (a) The atomic energy spectrum, weighted by the overlap of its state with the instantaneous eigenstates $|\braket{\psi(z,z_o)|\varphi_{\nu}(z,z_o)}|^2$ of Eq.~\eqref{eq:hamiltonian} as a function of $z_o(t)$. (b) Atomic probability density $|\psi(z,t)|^2$ throughout the protocol. The solid line is the harmonic trap potential (scaled for visibility) and the dashed line indicates $z_o(t)$. (c) Comparison of the density of the atom's final state $|\psi_f|^2$ to the target state $|\phi_3|^2$. Here, we achieve a fidelity $|\braket{\psi_f|\psi_{t}}|^2$ of 99.02\%. (d)-(f) Same as for the top row for the target state $(\phi_0+e^{i\Phi(t)}\phi_4)/\sqrt{2}$, where $\Phi(t) = -4\omega_z t$. Here, we achieve a fidelity $|\braket{\psi_f|\psi_{t}}|^2$ of 99.7\%. For both protocols, $v_{d} = 0.1$.} Wavefunctions are normalised such that $\int dz |\psi(z)|^2 = 1$.
    \label{fig:fig5}
\end{figure*}
\begin{table*}[t]
  \centering
  \begin{tabular}{|c|c|}
  \hline
    $\%$ error in $z_s$ & $|\braket{\psi_f|\phi_3}|^2$\\
    \hline
    0.01 & 98.70  \\ 
    0.10 & 72.14  \\ 
    1.00 & 1.18   \\ \bottomrule
    \hline
  \end{tabular}
  \caption[Robustness of tunnelling probability to error in stopping position.]{\textbf{Robustness of tunnelling probability to error in stopping position}. (I) Percentage error in stopping position $\bar{z}_s$. (II) Fidelity with the target state $\phi_3$. The error-free fidelity amounts to 99.02\% (see Fig.~\ref{fig:fig5}).}
  \label{table:tunnel-robustness}
\end{table*}
The key limiting factor of the protocols described in Section~\ref{sec:results-2} is their long duration: achieving fidelities with the target state greater than 90\% requires $10^3-10^4$ units of time, which translates to timescales on the order of seconds for trapping frequencies on the order of 100~Hz. Ideally, we want to be able to significantly reduce the duration of the protocols whilst still preserving their relative simplicity and high fidelity. This will be the focus of the following section. In Section~\ref{ssec:theory}, we show how more efficient protocols can be designed by drawing analogies between the dynamics of our system to the tunnelling of a particle in a double-well potential and arrive at a condition which enables tunnelling to be exploited usefully for state preparation in our system. In Section~\ref{ssec:results}, we apply the knowledge from Section~\ref{ssec:theory} to realise efficient protocols and present results for the preparation of pure and superposition excited trap states using the two varieties of avoided crossings in our system that were introduced in Section~\ref{sec:results-1}.\\
\subsection{Condition for complete tunnelling}\label{ssec:theory}
The combination of the harmonic trap and the dragged potential~\eqref{eq:potential} creates an effective potential for the atoms resembling an asymmetric double-well (cf. Fig.~\ref{fig:fig1}~(b) and~(c)). For the sake of building intuition, let us first consider the case of noninteracting atoms confined within a symmetric double-well potential, which is realised in our system for $z_o = 0$. The energy spectrum of atoms in a double well is characterised by a series of near-degenerate doublets whose eigenstates have opposite parity. Assume that at $t=0$ the atoms are in an equal superposition of the lowest two eigenstates: $\psi(z,0) = (\varphi_0(z) + \varphi_1(z))/\sqrt{2}$. From the near-degeneracy of the eigenstates $\varphi_0(z)$ and $\varphi_1(z)$ and their opposite parity, this wavepacket is localised solely within one of the wells. For $t>0$, the state undergoes unitary time evolution and accumulates a phase $\Phi$: $\psi(z,0) = (\varphi_0(z) + \exp{(i\Phi)}\varphi_1(z))/\sqrt{2}$, where $\Phi = -\Delta\varepsilon \,t$ which is proportional to the energy gap between the eigenstates $\Delta\varepsilon = \varepsilon_1-\varepsilon_0$. After a time $T=\pi/\Delta\varepsilon$, the state will have accumulated a phase $\pi$, such that the wavepacket is now localised within the opposite well: $\psi(T) = (\varphi_0 -\varphi_1)/\sqrt{2}$. For our purposes, we refer to $T$ as the tunnelling time.\\
Based on the size of the energy gaps at the avoided crossings in Fig.~\ref{fig:fig1}~(a), we can expect tunnelling times on the order of $10^2$ in our system. This value is one to two orders of magnitude smaller than the time required for the adiabatic protocols discussed in Section~\ref{sec:results-2} (see Fig.~\ref{fig:fig4} (c) and (h)). In other words, our estimate of the effective double-well tunnelling time $T$ for our system indicates that we could significantly lower the duration of our protocols by simply setting our adiabatic speed all the way to $v_a = 0$, i.e. stopping the potential in the vicinity of the avoided crossing and allowing the state to tunnel freely on timescales set by the atomic energy spectrum.\\
To exploit tunnelling for the purpose of state preparation, we need to understand how to control it. In this regard, two related questions arise. Firstly, what conditions must be fulfilled in the asymmetric double-well system to realise `perfect' tunnelling, namely where the atomic density tunnels completely from one side to the other without leaving behind any residue~? Secondly, can we realise such tunnelling for arbitrary positions of the dragged potential~? The remainder of this section provides concrete answers to these questions through some straightforward analytical considerations.\\
We assume that on the approach to the avoided crossing between the instantaneous eigenstates $\varphi_A$ and $\varphi_B$, the atomic state is in a superposition of only these two eigenstates: 
\begin{equation}\label{eq:wfn}
    \psi(z;z_o(t)) = c_A(z_o(t))\varphi_A(z;z_o(t)) + c_B(z_o(t))\varphi_B(z;z_o(t)),
\end{equation}
which is valid assuming that the dynamics up to this point has been diabatic. The complex coefficients $c_A(z_o(t))$ and $c_B(z_o(t))$ satisfy $|c_A(z_o(t))|^2 + |c_B(z_o(t))|^2 = 1$ since the atomic wavefunction is normalised $\braket{\psi(z;z_o(t))|\psi(z;z_o(t))}=1$. The narrowly-avoided crossing emerges due to a barrier created in the atoms' effective potential, centred at position $z_b$. Depending on the type of avoided crossing (see Section~\ref{sec:results-1} for details), $z_b$ may be equal to the position of the dragged potential $z_o(t)$, yet this is not guaranteed. For example, the variety of avoided crossings depicted in Fig.~\ref{fig:fig1}~(a) is not formed due to the external potential's Gaussian barrier but rather by its long-range attractive tail, hence in this case $z_b\neq z_o(t)$.\\
At $t=0$, the dragged external potential is suddenly halted at the position $z_o(0) = z_s$ near the avoided crossing between $\varphi_A$ and $\varphi_B$. Thereafter, the atomic wavefunction undergoes unitary evolution. Since the Hamiltonian $\hat{H}(z_s)$ no longer has explicit time-dependence, the wavefunction for $t\geq 0$ is given by $\psi(z,t;z_s) = e^{-i\hat{H}(z_s)t} \psi(z;z_s)$. In the interest of readability, we drop the $z_s$ parameter notation in equations beyond this point. The atomic probability density $\rho(z,t) = \psi^*(z,t)\psi(z,t)$ at time $t$ is given by:
\begin{equation}
    \label{eq:density-A}
    \begin{split}
        \rho(z,t) &= |c_A|^2|\varphi_A(z)|^2 + |c_B|^2|\varphi_B(z)|^2\\ 
        &+ 2c_Ac_B \cos({\Delta\varepsilon\, t})\varphi_A(z)\varphi_B(z),
    \end{split}
\end{equation}
where $\Delta\varepsilon$ is the energy difference between the eigenstates at position $z_s$ and we have assumed that the eigenstates are real-valued. For brevity, we label the time-independent and time-dependent contributions to the density as $\bar{\rho}(z) = |c_A|^2|\varphi_A(z)|^2 + |c_B|^2|\varphi_B(z)|^2 $ and $\delta\rho(z,t) = 2c_Ac_B \cos({\Delta\varepsilon\, t})\varphi_A(z)\varphi_B(z)$, respectively. Note that $\delta\rho(z,t)$ is periodic in time with period $P = 2\pi/\Delta\varepsilon$.\\
If the dynamics for $t < 0$ has been diabatic, the atoms' probability density at $t=0$ will be localised on one side of the barrier created in the effective potential, for example $z > z_b$ (see Fig.~\ref{fig:fig4}~(b)). Thus, the atomic density at $t=0$ fulfils the condition:
\begin{equation}
    \rho(z,0) = \bar{\rho}(z) + \delta\rho(z,0) = 0,\;\; \forall\; z \leq z_b.
\end{equation}
Using Eq.~\eqref{eq:density-A}, we can rewrite the above condition as:
\begin{equation}~\label{eq:condition-1}
    \bar{\rho}(z) = -\delta\rho(z,0) = - 2c_Ac_B \varphi_A(z)\varphi_B(z),\;\; \forall\; z \leq z_b.
\end{equation}
We now seek the optimal value of the external potential's stopping position, denoted by $\bar{z}_s$, such that the atoms undergo perfect tunnelling. This requires that at time $t=P/2$ the atoms are localised on the opposite side of the barrier in the effective potential. Hence, we demand that the atomic density fulfils the following condition:
\begin{equation}
    \rho(z,P/2) = \bar{\rho}(z) + \delta\rho(z,P/2) = 0,\;\; \forall\; z > z_b.
\end{equation}
Making use of Eq.~\eqref{eq:density-A} and $\cos({\Delta\varepsilon \, P/2}) = -1$ yields:
\begin{equation}\label{eq:condition-2}
        \bar{\rho}(z) = 2c_Ac_B\varphi_A(z)\varphi_B(z),\;\; \forall\; z > z_b.
\end{equation}
Finally, we make use of the conditions in Eq.~\eqref{eq:condition-1} and Eq.~\eqref{eq:condition-2} and the fact that $\bar{\rho}(z;z_s)$ is normalised to derive the following
\begin{equation}\label{eq:integrals}
\begin{split}
    1 = \int dz \, |\bar{\rho}(z)| &= \int_{z\leq z_b} dz \, |\bar{\rho}(z)| + \int_{z > z_b} dz \,|\bar{\rho}(z)| \\
    &=2|c_A||c_B|\int dz \,|\varphi_A(z)||\varphi_B(z)|\\
\end{split}
\end{equation}
In the above, we have used the absolute value in order to write the final expression as a single integral. Eq.~\eqref{eq:integrals} provides us with a relation between the overlap coefficients $c_i = \int dz \, \varphi_i(z)\psi(z)$ and the overlap of the eigenstates' absolute magnitudes $\mathcal{I} = \int dz |\varphi_A(z)||\varphi_B(z)|$ which must be fulfilled in order for perfect tunnelling to take place, namely  $|c_A||c_B|\,\mathcal{I}=1/2$.\\
Since $0\leq|c_A||c_B|\leq 1/2$ and $0\leq \mathcal{I} \leq 1$, the condition in Eq.~\eqref{eq:integrals} can only be fulfilled when $|c_A||c_B|=1/2$ and $\mathcal{I}=1$. This requires (i) the atomic state to be in an equal superposition of eigenstates $\varphi_A(z)$ and $\varphi_B(z)$ (i.e. $|c_A| = |c_B| = 1/\sqrt{2}$) and (ii) that these eigenstates differ at most by the sign of their prefactors ($|\varphi_A(z)|=|\varphi_B(z)| \;\;\forall\; z$). The former condition is rather loose, since it could be realised in general for arbitrary $z_s$. However, the latter condition provides a strong indication that the optimal stopping position $\bar{z}_s$ is located at the narrowest point of the avoided crossing between the eigenstates. Thus, we have shown that the requirements for perfect tunnelling in an asymmetric double well match those of the symmetric double well that we considered at the beginning of this section.
We determine $\bar{z}_s$ for a given crossing by evaluating the overlap integral of the eigenstates $|\varphi_A(z)|$ and $|\varphi_B(z)|$ for a range of $\bar{z}_s$ around their common avoided crossing. Fig.~\ref{fig:tunnel-condition} shows the results for $|\varphi_5(z)|$ and $|\varphi_6(z)|$. In this case, we confirm that the critical position $\bar{z}_s$ occurs at the point of closest approach between the energy curves $\varepsilon_5$ and $\varepsilon_6$.\\
In conclusion, the tunnelling protocol cannot be realised for arbitrary $\bar{z}_s$. In fact, the ability to tunnel is highly sensitive to the choice of $\bar{z}_s$ as shown by Fig.~\ref{fig:tunnel-condition}. Nonetheless, through the above analysis we have arrived at the condition $\mathcal{I}(\bar{z}_s) = 1$ which must be fulfilled to achieve perfect tunnelling, which provides us with a systematic method for determining the optimal stopping position $\bar{z}_s$. Furthermore, the size of the energy gaps at the avoided crossings mean that the atoms will tunnel over one order of magnitude faster than the adiabatic protocols discussed in Section~\ref{sec:results-2}.\\
\subsection{Proof-of-principle}\label{ssec:results}
Using the knowledge about the conditions for perfect tunnelling gained from the previous section, we now perform state preparation using with new protocols that execute sudden stops of the dragged potential at relevant avoided crossings in the atomic energy spectrum. The relevant avoided crossings are determined by the desired target state. The duration of each stop is set by the tunnelling time $T=\pi/\Delta\varepsilon$ for the given avoided crossing. Between stops, the external potential moves at a constant speed $\dot{z}_o=0.10$ and the change in its velocity is assumed to be sudden.\\
Fig.~\ref{fig:fig5} summarises the results of tunnelling protocols for target states $\psi_t^{(3)}(z) = \phi_3(z)$ and $\psi_t^{(4)}(z,t) = (\phi_0(z) + e^{i\Phi(t)}\phi_4(z))/\sqrt{2}$, where $\Phi(t) = -4\omega_z t$. In both cases, we achieve fidelities above 99\% for durations of $10^2$ time units. In order to prepare the superposition state $\psi_t^{(4)}(z,t)$, we follow a slightly different approach by exploiting instead the avoided crossings that arise between a bound state of the dragged potential~\eqref{eq:potential} and a vibrational state (see e.g. Fig.~\ref{fig:fig1} (b)). Using these anti-crossings requires us to reverse the direction of motion of the dragged potential, which therefore requires stopping twice during the protocol as compared to only once in the protocol on the top row of Fig.~\ref{fig:fig5}. The advantage of this approach is however that there are overall fewer avoided crossings that the state has to traverse, which improves the overall fidelity at the cost of a slightly longer protocol. Finally, we note that by stopping the potential at $\bar{z}_s$ for only half the tunnelling time $T/2$ the state will split equally along both paths that meet at the crossing. Using this method, we achieve a fidelity of 99.7\% with $\psi_t^{(4)}(z,t)$ in a time of 650 (see bottom row of Fig.~\ref{fig:fig5} for further details).\\
Whilst the tunnelling protocols have a distinct advantage in terms of speed, their major drawback is their sensitivity to errors in the stopping position $\bar{z}_s$. In Table~\ref{table:tunnel-robustness}, we summarise some data which investigates the robustness of the protocol to errors in the critical position $\bar{z}_s$. We find that deviations as small as 0.1\% can lead to a sizeable decrease in the fidelity with $\psi_t^{(3)}(z,t)$. The level of precision in the positioning of the potential might be challenging to meet by current experimental standards.\\
Additionally, the protocols discussed in this work will be sensitive to deviations in the swept potential from its optimal shape. This effect is investigated in Appendix~\ref{ssec:noise-robustness}.\\
In this work, we have demonstrated the preparation of superposition states consisting of at most two trap eigenstates in the population-balanced case. It is possible to modify the protocols to prepare more complicated superposition states, as discussed for the case of the tunnelling protocol in Appendix~\ref{ssec:three-state-prep}.\\
\section{Summary and conclusions}\label{sec:summary}
In this work, we explored protocols for exciting individual trapped atoms into higher vibrational states by means of a dynamically-swept external potential. In particular, we employed an external potential possessing long-range attractive character and a repulsive barrier at its centre, which could be realised via a tightly-trapped ion or a shaped optical potential. Excitation of the atoms was facilitated by avoided crossings in the atomic energy spectrum, whose position and gap size may be tuned through the shape of the external potential. The presence of the avoided crossings is a consequence of shape resonances in the effective potential created by the moving external potential, analogous to TISR emerging in collisions between species in separate trapping potentials. The protocols proposed in our work selectively prepare the atoms in excited vibrational states through controlling the speed of the external potential in order to drive the state along a desired path through the atoms' discrete energy spectrum.\\ 
The first protocol relies on adiabatic driving around a small number of critical anti-crossings, which depend on the desired target state. The protocol's primary limitation is its duration: achieving fidelities higher than 90\% requires durations of $10^3-10^4$ in harmonic oscillator units. For a Rb atom with $\omega_z = 2\pi\cdot1$~kHz, this would correspond to a protocol duration of approximately 0.1~s to 1.0~s.\\ 
In contrast, the second protocol brings the potential to a complete halt at the critical avoided crossings, whereupon the atom undergoes unitary dynamics in its effective potential created by the harmonic trap and the now static external potential. During this period, the atom tunnels through the barrier present at the shape resonance on timescales defined by the energy gap between the eigenstates at the avoided crossing. We found that tunnelling occurs over durations of $10^2$, which is one to two orders of  magnitude faster than the timescales for the adiabatic protocol. The tunnelling protocol achieved fidelities higher than 99\% with protocol durations of 10~ms to 100~ms, assuming a Rb atom with $\omega_z = 2\pi\cdot1$~kHz. However, the fidelity of this protocol is highly-sensitive to the external potential’s stopping position.\\
Without any specific attempts at optimisation, our protocols can achieve fidelities better than 99\% on timescales on the order of milliseconds. Whilst employing QOC methods would enable us to design protocols with more competitive durations, these protocols would not offer the same level of clarity and intuitiveness as the protocols presented in this work.\\
Our work may be extended to weakly-interacting Bose or Fermi gases to investigate the role of interparticle interactions and particle statistics. Moreover, considering a binary mixture may be of particular interest. For instance, consider a mixture of two components A and B, where species A initially occupies an excited trap state and species B occupies the vibrational ground state. Introducing weak interspecies interactions would mean that species B experiences, in an effective picture, a lattice-like background potential created by the density of species A. Additionally, the lattice could be made to vibrate by preparing species A in a superposition of trap states, thus mimicking phononic excitations.\\
Whilst preparing this work for submission, we became aware of earlier related works~\cite{Karkuszewski2001Method,Damski2001Simple} which propose preparing atoms in pure excited trap states through adiabatic passage of a constant-speed potential well with varying well depth.\\
\appendix
\section*{Acknowledgements}
This work is funded by the Cluster of Excellence “Advanced Imaging of Matter” of the Deutsche Forschungsgemeinschaft (DFG)-EXC 2056, Project ID No. 390715994.
%
\section*{Appendix}
\begin{figure}
    \centering
    \includegraphics[width = 0.5\textwidth]{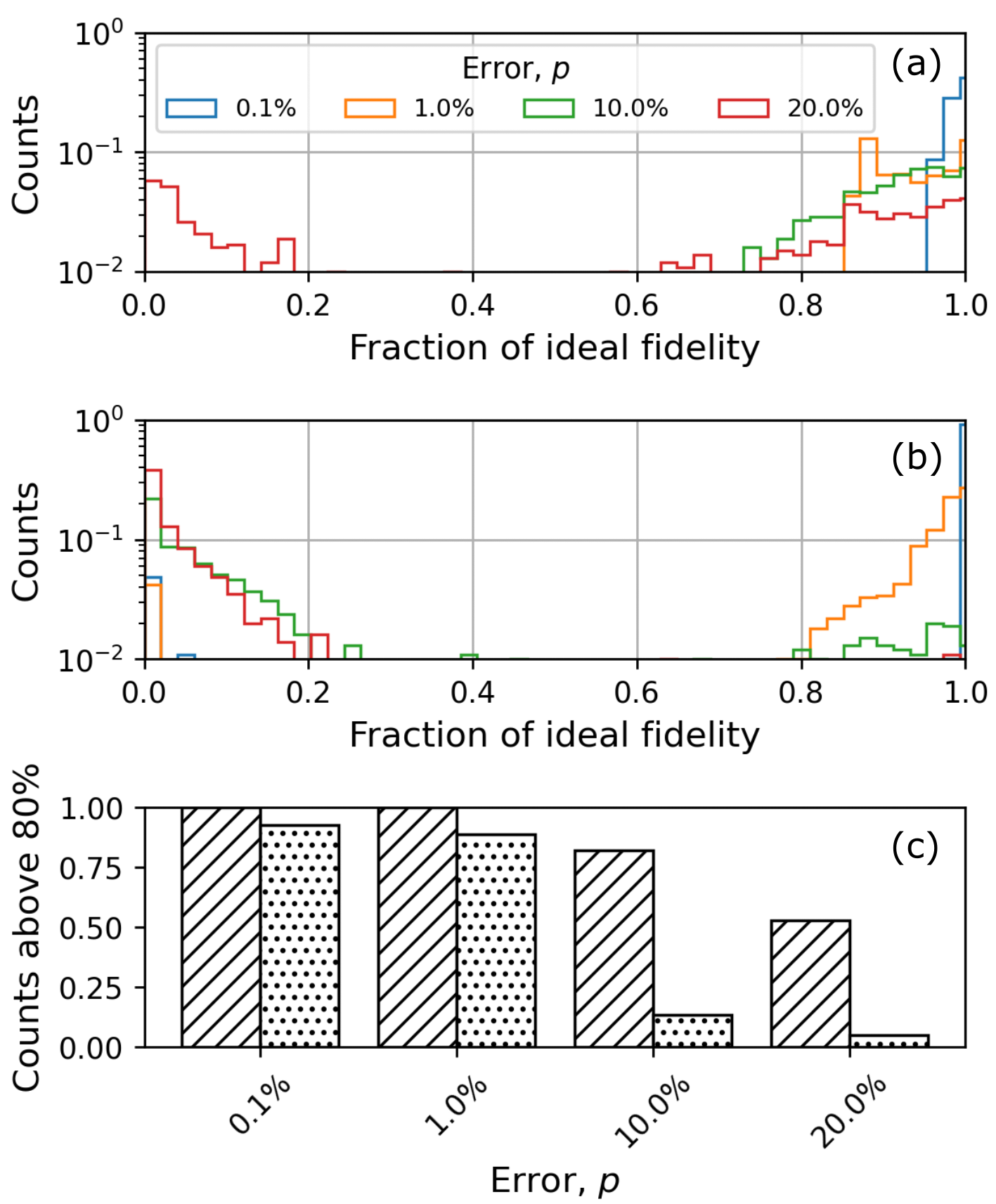}
    \caption{\textbf{Robustness of the protocols to deviation in shape of the swept potential}. (a) Distribution of fidelities achieved by the adiabatic protocol for varying degrees of error in the model parameters $p$ of the swept potential (see Eq.~\eqref{eq:potential}). For each value of $p$, one-thousand randomly-sampled swept potentials were used. The fidelities are plotted as a fraction of the ideal fidelity with respect to the target state and the counts are normalised. In this case, the target state was the fifth excited trap state $\phi_5$ with an ideal fidelity of 97.40\%. (b) Shows the corresponding results for the tunnelling state preparation protocol. In this case, the target state was the third excited trap state with an ideal fidelity of 99.02\%. (c) Number of simulations which achieved better than 80\% of the ideal fidelity for the adiabatic (slash-hatched) and tunnelling (dot-hatched) protocols.}
    \label{fig:noise}
\end{figure}
\subsection{Robustness of the protocols to errors in the swept potential}\label{ssec:noise-robustness}
The success of the protocols described in this paper in an experimental setting would rely on the ability to recreate the shaped potential~\eqref{eq:potential} with high precision. Deviations in the shape of the potential, through e.g. errors in the model parameters, will lead to a reduction in fidelity with respect to the target state. In this section, we investigate the robustness of our protocols to deviations in the potential's shape from the ideal, mimicking the impact of experimental errors.\\
We carried out a series of simulations using swept potentials whose model parameters ($a$, $b$ and $c$) were sampled from Gaussian distributions, where the mean was fixed at the ideal value ($a_0 = 120$, $b_0 = 4\sqrt{10c_0}$, $c_0 = 40$) and the standard deviation $\sigma$ was changed to reflect varying degrees of experimental error. Fig.~\ref{fig:noise} shows the distribution of fidelities achieved for the adiabatic and tunnelling protocols with varying standard deviation, defined as a percentage $p$ of the ideal value (i.e. the Gaussian distribution for model parameter $a$ would have a standard deviaton $\sigma = p a_0$). To obtain statistics, we performed simulations with one-thousand randomly-sampled swept potentials for each value of $p$.\\
As expected, the protocols perform worse for increasing noise. The tunnelling protocol is particularly sensitive: for $p=0.20$, fewer than 10\% of runs achieved a fidelity higher than 80\% of the ideal value. In contrast, the adiabatic protocol proved itself more robust, with around 50\% of simulations achieving a fidelity better than 80\% at $p=0.20$ (see Fig.~\ref{fig:noise}~(c).
\subsection{Further examples of state preparation}\label{ssec:three-state-prep}
\begin{figure*}[t]
    \centering
    \includegraphics[width=\textwidth]{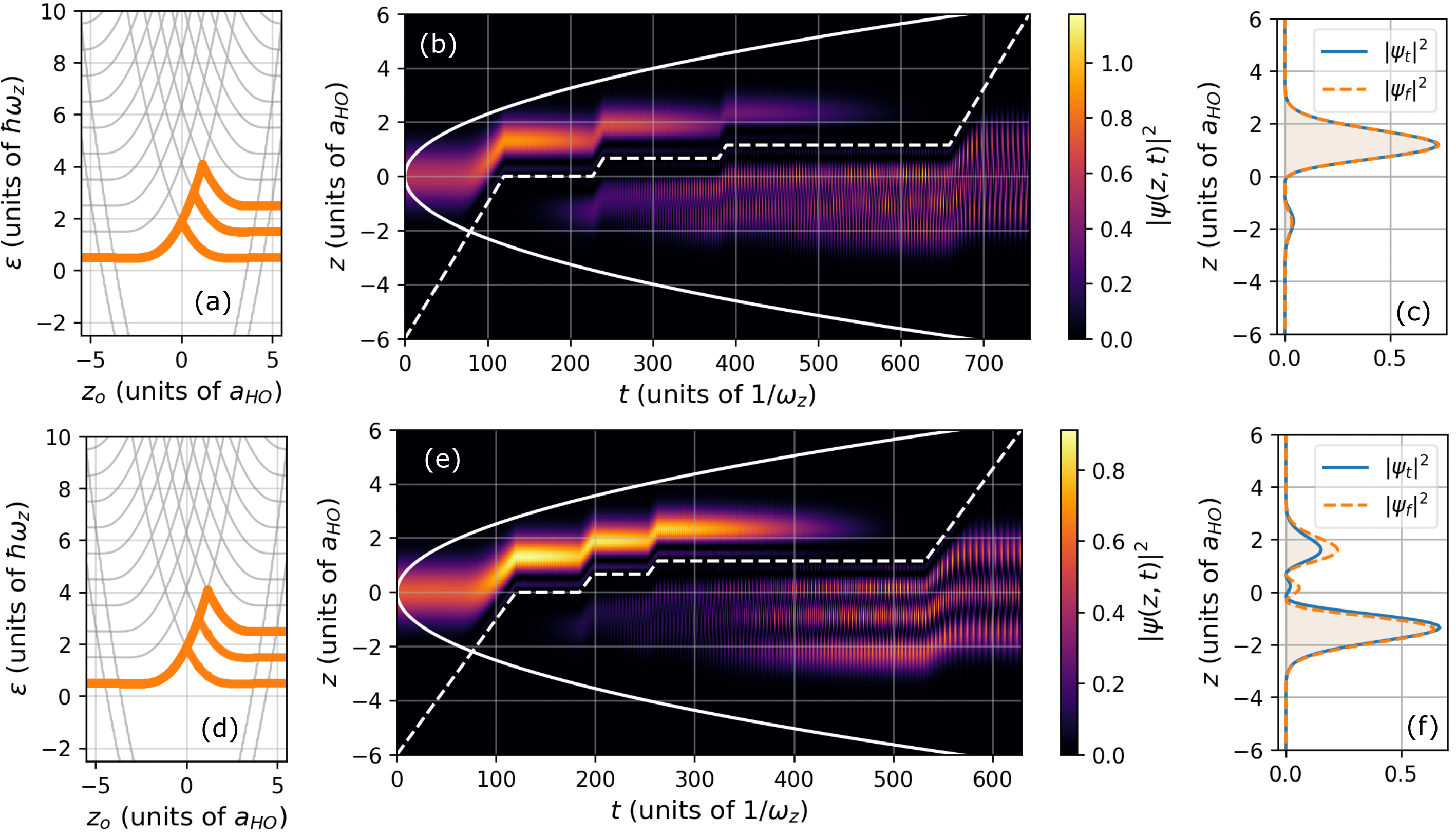}
    \caption[Tunnelling protocol results: proof-of-principle]{\textbf{Tunnelling protocol: three state superposition.} (a)-(c) Exciting atoms to the population-balanced three state mixture (see text) using the tunnelling protocol. (a) The atomic energy spectrum, weighted by the overlap of its state with the instantaneous eigenstates $|\braket{\psi(z,z_o)|\varphi_{\nu}(z,z_o)}|^2$ of Eq.~\eqref{eq:hamiltonian} as a function of $z_o(t)$. (b) Atomic probability density $|\psi(z,t)|^2$ throughout the protocol. The solid line is the harmonic trap potential (scaled for visibility) and the dashed line indicates $z_o(t)$. (c) Comparison of the density of the atom's final state $|\psi_f|^2$ to the target state $|\phi_t|^2$. Here, we achieve a fidelity $|\braket{\psi_f|\psi_{t}}|^2$ of 99.87\%. (d)-(f) Same as for the top row for the population-imbalanced three state mixture. Here, we achieve a fidelity $|\braket{\psi_f|\psi_{t}}|^2$ of 98.76\%. For both protocols, the speed of the ion between stops was $v_{d} = 0.1$. Wavefunctions are normalised such that $\int dz |\psi(z)|^2 = 1$.}
    \label{fig:three-state-prep}
\end{figure*}

In the main part of the manuscript, we show proof-of-principle results for the preparation of target states involving at most two excited trap states. Nonetheless, the protocols can be applied to realise more sophisticated superpositions of trap states. In this section, we show results for the preparation of a superposition of the three lowest-energy trap states:
\begin{equation}\label{eq:3-state}
    \psi_t = \alpha\phi_0 + \beta\phi_1 + \gamma\phi_2.
\end{equation}
Fig.~\ref{fig:three-state-prep} demonstrates results for the preparation of a population-balanced case ($\alpha=\beta=\gamma=1/\sqrt{3}$) as well as an imbalanced case with coefficients in the ratio $\alpha:\beta:\gamma = 5:4:16$ using the tunnelling protocol. We achieve fidelities of 99.87\% and 98.76\% for the two cases, respectively.\\
In general, even more complex superpositions of states could be created in this way. However, guiding the state along more complicated paths in the energy spectrum shown in Fig.~\ref{fig:fig1}~(a) would require the traversal of an increased number of avoided crossings which, in principle, means a larger overall loss of fidelity.\\

\end{document}